\documentstyle[12pt,epsfig]{article}
\def\lsim{\mathrel{\raise3pt\hbox to 8pt{\raise -6pt\hbox{$\sim$}\hss{$<$}}}}

\def\haf{\textstyle{1\over2}}
\def\thaf{\textstyle{3\over2}}

\def\minus{\mbox{$-$}}
\newcommand{\vr}{\vec{\, r}}
\newcommand{\vrp}{\vec{\,r}^{\prime}}
\newcommand{\vx}{\vec{x}}
\newcommand{\vxp}{\vec{x}^{\prime}}
\newcommand{\vy}{\vec{y}}
\newcommand{\vz}{\vec{z}}

\newcommand{\vq}{\vec{\, q}}
\newcommand{\vD}{\vec{D}}
\newcommand{\vJ}{\vec{J}}
\newcommand{\vL}{\vec{L}}
\newcommand{\vS}{\vec{S}}
\newcommand{\vsig}{\vec{\sigma}}
\newcommand{\valf}{\vec{\alpha}}
\newcommand{\vmu}{\vec{\mu}}
\newcommand{\muN}{\mu^{\phantom{|}}_N}
\newcommand{\mun}{\mu^{\phantom{|}}_n}
\newcommand{\mup}{\mu^{\phantom{|}}_p}
\newcommand{\mud}{\mu^{\phantom{|}}_d}
\newcommand{\muH}{\mu^{\phantom{.}}_{^3{\! \rm H}}}
\newcommand{\muHe}{\mu^{\phantom{.}}_{^3{\! \rm He}}}
\newcommand{\vpi}{\vec{\pi}}
\newcommand{\vnabla}{\vec{\nabla}}

\newcommand{\he}{\hat{e}}
\newcommand{\hmu}{\hat{\mu}}
\newcommand{\hp}{\hat{p}}
\newcommand{\hn}{\hat{n}}
\newcommand{\hr}{\hat{r}}
\newcommand{\hz}{\hat{z}}
\newcommand{\hxij}{\hat{x}_{i j}}
\newcommand{\bC}{\bar{C}}
\newcommand{\bbC}{\bar{\! \bar{C}}}
\newskip\humongous \humongous=0pt plus 1000pt minus 1000pt
\def\caja{\mathsurround=0pt}

\newif\ifdtup
\def\panorama{\global\dtuptrue \openup1\jot \caja
        \everycr{\noalign{\ifdtup \global\dtupfalse
        \vskip-\lineskiplimit \vskip\normallineskiplimit
        \else \penalty\interdisplaylinepenalty \fi}}}
\def\eqalignno#1{\panorama \tabskip=\humongous
        \halign to\displaywidth{\hfil$\displaystyle{##}$
        \tabskip=0pt&$\displaystyle{{}##}$\hfil
        \tabskip=\humongous&\llap{$##$}\tabskip=0pt
        \crcr#1\crcr}}
\begin{document}
\vspace*{-0.6in}
\hfill \fbox{\parbox[t]{1.12in}{LA-UR-04-5063}}\hspace*{0.35in}
\vspace*{0.0in}

\begin{center}

{\Large {\bf Nuclear Corrections to Hyperfine Structure\\
in Light Hydrogenic Atoms}}\\

\vspace*{0.4in}
{\bf J.\ L.\ Friar} \\
{\it Theoretical Division,
Los Alamos National Laboratory \\
Los Alamos, NM  87545} \\
\vspace*{0.10in}
\vspace*{0.10in}
and \\
\vspace*{0.10in}
{\bf G.\ L.\ Payne}\\
{\it Dept. of Physics and Astronomy\\
Univ. of Iowa\\
Iowa City, IA 52242}
\end{center}

\begin{abstract}
Hyperfine intervals in light hydrogenic atoms and ions are among the most
accurately measured quantities in physics. The theory of QED corrections has
recently advanced to the point that uncalculated terms for hydrogenic atoms and
ions are probably smaller than 0.1 parts per million (ppm), and the experiments
are even more accurate. The difference of the experiments and QED theory is
interpreted as the effect on the hyperfine interaction of the (finite) nuclear
charge and magnetization distributions, and this difference varies from tens to
hundreds of ppm. We have calculated the dominant component of the 1s hyperfine
interval for deuterium, tritium and singly ionized helium, using modern
second-generation potentials to compute the nuclear component of the hyperfine
splitting for the deuteron and the trinucleon systems. The calculated nuclear
corrections are within 3\% of the experimental values for deuterium and tritium,
but are about 20\% discrepant for singly ionized helium. The nuclear corrections
for the trinucleon systems can be qualitatively understood by invoking SU(4)
symmetry.
\end{abstract}
\pagebreak
\section{Introduction}

The physics of hyperfine structure (hfs) is driven by magnetic interactions. 
This physics has a short-range nature, and is more complicated and challenging
than ``softer'' regimes in atomic physics.  This is especially true of the
nuclear contribution to hyperfine structure, because the nuclear current density
is more complicated than the nuclear charge density and less well
understood\cite{MEC,primer}.

Until very recently hyperfine splittings in light hydrogenic atoms were by far
the most precisely measured atomic transitions.  Many of these very accurate
experiments date back nearly half a century.  Theoretical predictions are far
less accurate, but have improved considerably in recent years.  Non-recoil and
non-nuclear contributions\cite{eides01,savely02} are known through order
$\alpha^3 E_{\rm F}$, where $E_{\rm F}$ is the Fermi hyperfine energy (viz., the
leading-order contribution) and $\alpha$ is the fine-structure constant. 
Because the hadronic scales for recoil and certain types of nuclear corrections
are the same, recoil corrections are treated on the same footing as nuclear
corrections\cite{eides01}, and we will call both types ``nuclear corrections.'' 
It is very likely that the uncalculated QED terms of order $\alpha^4 E_{\rm F}$
in light atoms contribute less than 0.1 ppm. Although much larger than the
experimental errors, this is still significantly smaller than the nuclear
corrections. We restrict ourselves to hydrogenic s-states in this work, because
these states maximize nuclear effects.

\begin{table}[htb]
\centering
{\bf Table I}.
\caption{Difference between hyperfine experiments and QED hyperfine calculations
for the $n\underline{\rm th}$ s-state of light hydrogenic atoms times $n^3$,
expressed as parts per million of the Fermi energy. This difference is
interpreted as nuclear contributions to the hyperfine
splitting\protect\cite{savely02}. A negative entry indicates that the
theoretical prediction without nuclear corrections is too large}
\vspace*{0.1in}
\begin{tabular}{|l || c c c c|}
\multicolumn{5}{c}{$n^3 (E_{\rm hfs}^{\rm exp} - 
E_{\rm hfs}^{\rm QED})/E_{\rm F}\, {\rm (ppm)}$} \\ \noalign{\smallskip} \hline
 State & H & $^2$H & $^3$H & $^3$He$^+$ \rule{0in}{2.5ex}\\ \hline \hline
 1s    & $-$33   & 138    & $-$38       & $-$212        \\ \hline
 2s    & $-$33   & 137    &\mbox{$--$}  & $-$211        \\ \hline 
\end{tabular}
\end{table}

Table I is an updated version of the corresponding table in Ref.\cite{savely02}.
Because hyperfine splittings are dominated by short-range physics, we expect the
splittings in the $n\underline{th}$ s-state to be proportional to
$|\phi_n(0)|^2 \sim 1/n^3$, where $\phi_n (r)$ is the non-relativistic wave
function of the electron. Forming the fractional differences (in parts per
million) between $E^{\rm exp}_{\rm hfs}$ and $E^{\rm QED}_{\rm hfs}$ leads to
the tabulated results. These large differences reflect neither experimental
errors nor uncertainties in the QED calculations; they directly reflect large
nuclear contributions.

One complication in performing the nuclear calculation is obtaining a final
result that is tractable for numerical calculations.  A framework fortunately
exists for performing systematic expansions\cite{primer} of nuclear matrix
elements in powers of $(Q/\Lambda)$, where $Q$ is a typical nuclear momentum
scale that can be taken to be roughly the pion mass ($m_\pi \sim$ 140 MeV), and
$\Lambda$ is the large-mass QCD scale ($\sim$ 1 GeV) typical of QCD bound states
such as the nucleon, heavy mesons, nucleon resonances, etc.  This framework,
called power counting, also extends to nuclei, where $1/Q$ specifies a typical
correlation length (and a reasonable nearest-neighbor distance) in light nuclei
($\sim$ 1.4 fm).  This expansion in powers of $(Q/\Lambda \sim 0.1-0.15)$ should
converge moderately well.  For the purposes of this initial work, we restrict
ourselves to leading-order terms in the nuclear corrections.

This restriction eliminates nuclear corrections of relativistic order, which we
will briefly discuss later. There have been relatively few relativistic
calculations in light nuclei because of the complexity of the nuclear force. The
calculations that exist are known to generate rather small
corrections\cite{tjon}, and they are especially small in the deuteron case
because of its weak binding\cite{chi-pt}. Few calculations exist for the much
more complicated three-nucleon systems\cite{franz}. Most nuclear physics
knowledge and lore in light nuclei is non-relativistic in nature.

In processes that involve virtual excitation of intermediate nuclear states
(each state $| N \rangle$ with its own energy, $E_N$, relative to the
ground-state energy, $E_0$) the excitation energy $(\omega_N = E_N - E_0)$ is of
order $Q^2/\Lambda$ and typically is a correction to the leading
order\cite{primer}. Consistency therefore demands that we drop such terms.  The
nuclear recoil energy $Q^2/2 M$, where $M$ is the nuclear mass, has the same
scale and can also be dropped.

Our goal is to evaluate the deuteron contribution in leading order as carefully
as possible, and to use the impulse approximation to evaluate the $^3$He and
$^3$H results. This restriction should be accurate to within the 10-15\%
uncertainty of our leading-order approximation and adequate for a first
calculation of the trinucleon sector.  This level of accuracy does require the
inclusion of the intrinsic hyperfine structure of the nucleons. We will see that
the calculated results are good for deuterium and tritium and fairly good for
$^3$He$^+$. Our goal is to present a simple and compelling picture of the
nuclear hyperfine structure in $d$, $^3$H, and $^3$He$^+$, using a unified
approach for all. To accomplish this we invoke a simple and intuitive model of
trinucleon structure based on SU(4) symmetry that is sufficiently accurate to
explain the patterns in Table~I.

\section{Nuclear Contributions to Hyperfine Structure}

The hyperfine interactions that interest us are simple (effective) couplings of
the electron spin to the nuclear (ground-state) spin: $\vsig \cdot \vS$ where
$\vsig$ is the electron (Pauli) spin operator and $\vS$ is the nuclear spin
(i.e., the total angular momentum\cite{edmonds}) operator. Other couplings of
the electron spin are possible and either generate no hyperfine splitting, none
in s-states, or higher-order (in $\alpha$) contributions. Coupling of the
electron spin to the electron angular momentum (i.e., the electron spin-orbit 
interaction, which vanishes in s-states) is one example.

We begin with a sketch of the Fermi hyperfine splitting in order to establish
our notation and conventions.  The electron charge operator ($\psi^{\dagger}
\psi$) is a space scalar, and by itself does not generate a term of the desired
form because nuclear-spin information cannot be transmitted via the exchange of
the appropriate component of the virtual photon's propagator. The leading-order
term in the hyperfine energy shift consequently has the (usual) form of a
current-current interaction.
$$
E_{\rm F} = \alpha \, \int d^3 r \int d^3 r^{\prime} \, \frac{\psi^{\dagger} 
(\vr) \valf \, \psi (\vr)\, \cdot \vJ (\vrp)}{|\vr-\vrp|}  \, , \eqno(1)
$$
where $\psi^{\dagger} \valf \psi$ is the current density of the (Dirac) electron
and $\vJ (\vrp)$ is the nuclear (ground-state) current density.  Ignoring
higher-order (in $\alpha$) terms this expression can be manipulated into the 
form
$$
E_{\rm F} = \frac{\alpha}{2 m_e}\int d^3 r \; |\phi_n (r)|^2
\int d^3 r^{\prime} \, \vJ (\vrp) \cdot \vsig \times \vnabla_r 
\frac{1}{|\vr-\vrp|}  \, , \eqno(2)
$$
where $m_e$ is the electron mass and $\phi_n (r)$ is the usual non-relativistic
atomic $n\underline{\rm th}$ s-state wave function. The restriction to atomic
s-states limits the current to the magnetic-dipole part, which can be written in
terms of the magnetic-dipole operator, $\vmu$, as
$$
\vJ (\vr) \rightarrow \vJ_{M1} (\vr) = \vnabla_r \times \vmu (\vr) \, , 
\eqno(3a)
$$
and $\vmu$ can be written in terms of the nuclear ground-state magnetization 
density $\rho_M (r)$ as
$$
\vmu (\vr) =  \frac{\muN\, \vS}{S}\, \rho_M (r) \, , \eqno(3b)
$$
where $\int d^3\, r^{\prime} \rho_M (r^{\prime}) = 1$ and $\langle S S |
\mu_{\rm z} |S S \rangle \equiv \muN$ defines the nuclear magnetic moment.  We
finally obtain the Fermi hyperfine energy in the limit $\rho_M (r^{\prime})
\rightarrow \delta^3 (\vrp)$:
$$
E_{\rm F} = \frac{4 \pi \alpha \muN}{3 m_e}\frac{\vsig \cdot \vS}{S}
\int d^3 r |\phi_n (r)|^2 \rho_M (r) = \frac{4 \pi \alpha \muN |\phi_n (0)|^2}
{3 m_e}\frac{\vsig \cdot \vS}{S} \, , \eqno(4)
$$
where the factor of $(\vsig \cdot \vS)/S$ leads to a hyperfine splitting of
$(2\, S + 1)/S$. All additional contributions will be measured as a fraction of
this energy.  We note that using the point-like $\rho_M$ and Eqn.~(3a) in
Eqn.~(1) leads to an atomic matrix element $\int_0^{\infty} dr F(r) G(r)$ (in
the standard Dirac notation) that can be evaluated analytically for higher-order
Coulomb corrections\cite{eides01} to the leading contribution of ${\cal O} (Z^3
\alpha^4)$, where $Z$ is the nuclear charge. Note that $E_{\rm F}$ is
independent of nuclear structure.

\begin{figure}[htb]
\epsfig{file=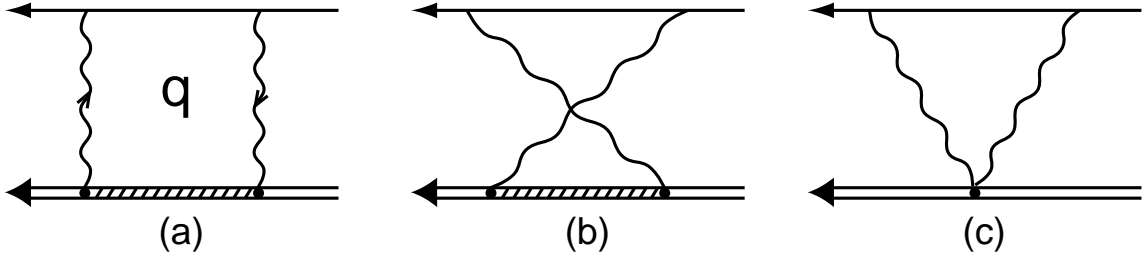,height=1.32in}
\caption{Nuclear Compton amplitude with direct (a), crossed (b), and seagull (c)
contributions illustrated. Single lines represent an electron, double lines a
nucleus, wiggly lines a photon propagator (with four-momentum $q^{\mu}$), and 
shaded double lines depict a nuclear Green's function containing a sum over
nuclear states. The seagull vertex maintains gauge invariance.}
\end{figure}

Naively calculating higher-order (in $\alpha$) corrections from the first part
of Eqn.~(4) fails because only that part of $\phi_n (r)$ inside the nucleus
(i.e., within the magnetic ``size'', $R_M$) contributes to the integral, and
that is the (only) part of the electron's wave function significantly modified
by the nuclear charge distribution, $\rho_{\rm ch} (r)$, whose radius, $R_{\rm
ch}$, is nearly the same as $R_M$.  In other words a proper
calculation\cite{BW,zemach56} must take into account modifications of $\phi_n
(r)$ by $\rho_{\rm ch}$, and that necessarily involves one order in $\alpha$
higher than $E_{\rm F}$.  Thus we need to perform a consistent second-order (in
$\alpha$) calculation of the electron-nucleus interaction, depicted in Fig.~(1).

 These graphs are constructed from the direct, crossed, and nuclear seagull
contributions to the nuclear Compton amplitude.  Only the forward-scattering
part of this amplitude is required for the ${\cal O} (\alpha)$ corrections to
$E_{\rm F}$, and this generates a short-range atomic operator that samples the
(upper-component) s-state wave functions only near the origin. The resulting
energy shift is then given by
$$
\Delta E = i (4 \pi \alpha)^2 |\phi_n (0)|^2 \int \frac{d^4 q}{(2 \pi)^4}
\frac{t_{\mu \nu} (q)\, T^{\mu \nu} (q,-q)}
{(q^2 + i \epsilon)^2 (q^2 - 2 m_e q_0 + i \epsilon)}\, , \eqno(5)
$$
where $t^{\mu \nu}$ is the lepton Compton amplitude and $T^{\mu \nu}$ is the 
corresponding nuclear Compton tensor, both of which are required to be gauge 
invariant.  The lepton tensor can be decomposed into an irreducible spinor
basis; we can ignore odd matrices and spin-independent components, since they do
not contribute to the hyperfine structure at ${\cal O} (\alpha E_{\rm F})$. We
also ignore (with one exception treated later) terms that couple two currents
together. It is easy to show that since the nuclear current scales as
$1/\Lambda$ (the conventional components of the current have explicit factors of
$1/M$), two of them should scale as $1/\Lambda^2$ and generate higher-order (in
$1/\Lambda$) terms.  This leaves a single term representing a charge-current
correlation
$$
\Delta E = (4 \pi \alpha)^2 |\phi_n (0)|^2 \int \frac{d^4 q}{(2 \pi)^4}
\frac{(\vsig \times \vq)^m [T^{m 0} (q,-q) - T^{0 m} (q,-q)]}
{(q^2 + i \epsilon)^2 (q^2 - 2 m_e q_0 + i \epsilon)}\, . \eqno(6)
$$
The seagull terms $B^{0 m}(q,-q)$ and $B^{m 0}(q,-q)$ are of relativistic
order\cite{dhg} $(\sim 1/\Lambda^2)$ and can be dropped. Although the term $B^{m
n} (\vq,-\vq)$ is of non-relativistic order, it does not contribute in
conventional approaches because of crossing symmetry. 

The explicit form for $T^{m 0}$ (suppressing the nuclear ground-state 
expectation value, but including all intermediate states, $N$) is
$$
T^{m 0} (q,-q) = \sum_{N} \left( 
\frac{\, J^m (-\vq)\, |N \rangle \langle N | \, \rho (\vq)}
{q_0 - \omega_N + i\epsilon} + 
\frac{\, \rho (\vq)\, |N \rangle \langle N | \, J^m (-\vq)}
{- q_0 - \omega_N + i\epsilon} \right)
\, , \eqno(7)
$$
which greatly simplifies Eqn.~(6) in the limit $\omega_N \rightarrow 0$ and $m_e
\rightarrow 0$, leading finally in this limit to a very simple form when closure
is used
$$
\Delta E = i (4 \pi \alpha)^2 |\phi_n (0)|^2 \int \frac{d^3 q}{(2 \pi)^3}
\frac{(\vsig \times \vq)^m \{J^m (-\vq) , \rho (\vq)\}}
{\vq^6}\, , \eqno(8)
$$
which is infrared divergent.  Using
$$
J^m (-\vq) = \int d^3 y\, J^m (\vy) e^{-i \vq \cdot \vy} \, , \eqno(9a)
$$
and
$$
\rho (\vq) = \int d^3 x \, \rho (\vx) e^{i \vq \cdot \vx}\, , \eqno(9b)
$$
together with $\vz \equiv \vx - \vy$, and a lower-limit (infrared) $q-$cutoff, 
$\epsilon$, we find
$$
\Delta E = -8 \alpha^2 |\phi_n (0)|^2 \int d^3 x \int d^3 y \, \{ \rho (\vx), 
\vsig \cdot \vJ (\vy) \} \times \vnabla_{\rm z} \left(\frac{1}{3 \epsilon^3} - 
\frac{z^2}{6 \epsilon} +\frac{\pi z^3}{48}\right) \, , \eqno(10)
$$
where there is an implied (nuclear) expectation value. The constant term does
not contribute because of the derivative, the $z^2$-term contributes a term
proportional to $\frac{-(\vx-\vy)}{3 \epsilon}$, and the last term is the term
we are seeking.

The remaining singular (second) term must be treated more carefully.  The part
proportional to $\vy$ leads (because of the integral over $\vx$) to a
contribution proportional to $Z$, the total nuclear charge, and $\muN$, the
nuclear magnetic moment.  Since this contribution is already part of $E_{\rm
F}$, keeping this term would amount to double counting, and we therefore ignore
it.  The $\vx$-term on the other hand generates (unretarded) dipole transitions,
and the singularity $(1/\epsilon)$ arises from neglecting $1/\omega_N$ and $m_e$
terms. Siegert's theorem\cite{MEC} for unretarded electric dipole transitions ($
\int d^3 y \vJ (\vy) \sim i \omega_N \vD$) generates an additional factor of
$\omega_N$ (i.e., $\epsilon$) so this singular term is actually of the form
$(\epsilon/\epsilon) \rightarrow (0/0)$ and requires a careful calculation.  A
rather tedious evaluation of this term in Eqn.~(6) leads to $\Delta E_a = -2i
\alpha^2 |\phi_n (0)|^2 \vsig \cdot \vD$ $(\ln \frac{2(H-E_0)}{m_e} + 3/2) 
\times \vD$ where $H-E_0 \equiv \omega_N$ in intermediate states.  A similar 
term arises in the product of two currents mentioned above Eqn.~(6) and leads 
to $\Delta E_b = -i \alpha^2 |\phi_n(0)|^2 \vsig \cdot \vD$ $(\ln
\frac{2(H-E_0)}{m_e} - 1/2) \times \vD$.

These contributions are completely unimportant, as we now demonstrate.  The
constant terms (including $\ln (m_e)$) are proportional to $\vsig \cdot
\vec{D} \times \vec{D}$, which vanishes for non-relativistic dipole operators
because they commute.  Replacing $\omega_N$ by a constant (viz., the closure
approximation) similarly vanishes.  It is straightforward to show that the
nuclear matrix element also vanishes in zero-range approximation, where one
neglects the deuteron d-state and the potential in intermediate states (the
structure of the dipole operator weights the tails of the wave functions, and
this minimizes the effect of the intermediate-state nuclear potential).  In
perturbation theory it is possible to show that only the spin-orbit combination
of potentials in intermediate states contributes (i.e., central and tensor terms
cancel), and this small potential is of relativistic order $(1/\Lambda^2)$,
which we have previously agreed to ignore.  One can also show that the
non-vanishing deuteron contribution is proportional to $\eta^2$, the square of
the d- to s-wave asymptotic normalization constant $(\sim (0.025)^2)$, which is
extremely small.  A recent brute-force numerical calculation confirms these
estimates\cite{dipole}. The terms $\Delta E_a$ and $\Delta E_b$ are therefore
numerically negligible and can be ignored.

Our final result in leading order is a relatively simple expression originally
developed in a limiting case by Low\cite{d-th-1} for the deuteron, as sketched 
by Bohr\cite{d-th-0} for the same system:
$$
\Delta E^{\rm Low}_{\rm hfs} = \frac{\pi \alpha^2 |\phi_n (0)|^2}{2} 
\int d^3 x \int d^3 y \, \{ \rho (\vx), \vsig \cdot \vz \times \vJ (\vy) \} \, z
+ \cdots \, , \eqno(11a)
$$
where both an atomic and nuclear expectation value is implied, but has been 
ignored in Eqn.~(11a) and subsequent equations for reasons of simplicity.

A more convenient representation of this result is obtained by dividing both
sides by the expression for the Fermi hyperfine energy given by Eqn.~(4).  Since
the Wigner-Eckart Theorem guarantees that the resulting form of Eqn.~(11a) must 
be proportional to $\vsig \cdot \vS/S$ (which cancels in the ratio), we arrive 
at a simple but powerful expression for the leading-order contribution:
$$
\Delta E^{\rm Low}_{\rm hfs} = -2 m_e \, \alpha \, \delta_{\rm Low}\, , 
\eqno(11b)
$$
where
$$
\delta_{\rm Low} = -\frac{3}{16 \muN} \int d^3 x \int d^3 y \, \{ \rho (\vx),
(\vz \times \vJ (\vy))_{\rm z} \} | \vx-\vy | \, , \eqno(11c)
$$
and an expectation value is required of the z (or ``3'') component of the
vector $\vz \times \vJ (\vy )$ in the nuclear state with maximum azimuthal 
spin (i.e. $S_{\rm z} = S$). The intrinsic size of the nuclear corrections is
given by ($-2\, m_e\, \alpha R$) =  $-$38 ppm $[R/$fm], where $[R/$fm] is the
value of the Low moment in Eqn.~(11c) in units of fm. The results of Table~I
therefore suggest (correctly) that Low moments are on the order of a few fermi
in light nuclei, which is quite sensible.

\section{Nuclear Matrix Elements}

We predicate our discussion with the deuteron in mind.  Other nuclei can and
will be treated {\it mutatis mutandis}.  The isospin of the deuteron ($T = 0$)
makes it a useful first case.  We note that the nuclear physics in Eqn.~(11)
involves the correlation between the nuclear charge operator, $\rho (\vx)$, and
the nuclear current operator, $\vJ(\vy)$.  If one inserts a complete set of
states between these operators, there will be both elastic contributions (i.e.,
ground-state expectation values) that are called Zemach
corrections\cite{zemach56}, and inelastic contributions (called nuclear
polarization corrections). Although we will calculate (or estimate) both types,
it is much easier to calculate the sum of the two.

The nuclear charge operator contains both isoscalar and isovector pieces, and is
non-relativistic in leading order.  We ignore relativistic corrections, as we
discussed earlier.  We therefore write the charge operator in the form
$$
\rho (\vx) = \sum_{i = 1}^{A} \hat{e}_i (|\vx -\vxp_i|) \, , \eqno(12a)
$$
where
$$
\hat{e}_i (|\vx -\vxp_i|) = \hp_i \, \rho^p_{ch} (|\vx -\vxp_i|) 
+ \hn_i \, \rho^n_{ch} (|\vx -\vxp_i|) \, , \eqno(12b)
$$
and
$$
\hp_i = \left(\frac{1 + \tau^3_i}{2}\right) \, , \eqno(12c)
$$
and
$$
\hn_i = \left(\frac{1 - \tau^3_i}{2}\right) \, . \eqno(12d)
$$
This decomposes the i$\underline{th}$ nucleon's charge operator into proton plus
neutron parts. The densities $\rho^p_{ch}$ and $\rho^n_{ch}$ are the intrinsic
charge densities of the proton and neutron, respectively, while $\hp_i$ and
$\hn_i$ are the proton and neutron isotopic projection operators, respectively.
The coordinate $\vxp_i$ is the distance of the i$\underline{th}$ nucleon from
the nuclear center of mass. We expect that the neutron charge density should
play a very minor role, and we will find (later) that its contribution is only a
few percent of that of the proton.  Rather sophisticated models exist for the
Fourier transform (i.e., the form factor) of $\rho^p_{ch}$\cite{ingo03}.

The nuclear current operator is more complicated, even if we ignore relativistic
corrections (which we will). The problem is the mechanism underlying the nuclear
force (viz., the exchange of charged mesons), which can also contribute to the
nuclear current in the form of meson-exchange currents (MEC). These currents 
largely vanish for isoscalar transitions (such as the deuteron ground state)
because there is no net flow of charge, but they can be sizable (10\% - 20\%) 
for isovector transitions.  One can show that their contribution to the deuteron
in Eqn.~(11) almost entirely vanishes, and we will henceforth ignore these
currents below Eqn.~(13).  We formally expand the current into convection,
spin-magnetization, and meson-exchange parts
$$
\vJ (\vy) = \vJ_C (\vy) + \vnabla_y \times \vmu (\vy) + \vJ_{\rm MEC} (\vy)
+ \cdots \, , \eqno(13a)
$$
where
$$
\vJ_C (\vy) = \sum_{i=1}^A \left\{ \frac{\vpi_i}{2 M} , \hat{e}_i (|\vy-\vxp_i|)
\right\} \,  \eqno(13b)
$$
is the nuclear convection current, $\vpi_i$ is the i$\underline{th}$ nucleon's
momentum in the nuclear center-of-mass frame,
$$
\vmu (\vy) = \sum_{i=1}^A \vsig_i \, \hat{\mu}_i (|\vy-\vxp_i|) \,  \eqno(13c)
$$
is the impulse-approximation magnetic-moment operator, $\vJ_{\rm MEC}$ is the 
nuclear meson-exchange current, and
$$
\hat{\mu}_i (|\vy-\vxp_i|) = \hp_i \, \mup \, \rho^p_M (|\vy -\vxp_i|) 
+ \hn_i \, \mun \, \rho^n_M (|\vy -\vxp_i|) \, \eqno(13d)
$$
is the nucleon magnetization density for the i\underline{th} nucleon expressed
in terms of protons and neutrons separately.  The quantities $\mup$ and $\mun$
are the (total) proton and neutron magnetic moments, while $\rho^p_M$ and
$\rho^n_M$ are the intrinsic proton and neutron magnetization densities
(normalized to 1).

Using Eqns.~(12) and (13) the energy shift in Eqns.~(11) can be evaluated. 
Rather than split Eqn.~(11c) into Zemach terms (by inserting intermediate ground
states between $\rho$ and $\vJ$) and polarization terms (by inserting
intermediate excited nuclear states between $\rho$ and $\vJ$), we will use the
fact that our charge and current operators are each given by a sum over
single-nucleon operators. Thus their product can be decomposed into
single-nucleon plus two-nucleon operators. These forms are particularly
convenient to evaluate.  We first write in an obvious notation that
$$
\delta_{\rm Low} = \delta_{\rm spin}^{(1)} + \delta_{\rm Low}^{(2)} \, .
\eqno(14)
$$
Note that the quantity $\delta_{\rm spin}^{(1)}$ was not part of Low's original
work, nor was there any evidence at that time that such a term might be
significant. We next use Eqns.~(13) to manipulate the magnetization part of the
current in Eqn.~(11c) into the form (recall that $\vz = \vx - \vy$)
$$
\delta_{\rm Low}^{\rm mag} = \frac{1}{\muN}\sum_{i,j}^A 
\int d^3 x \int d^3 y \, |\vx-\vy| \, \hat{e}_i (|\vx-\vxp_i|) \hat{\mu}_j 
(|\vy-\vxp_j|) \left( \vsig_j - \frac{1}{8}(3 \vsig_j \cdot \hz \hz 
-\vsig_j)\right)_{\rm z} \, , \eqno(15)
$$
while the convection current can be reduced to
$$
\delta_{\rm Low}^{\rm conv} = -\frac{3}{16 \muN M}\sum_{i,j}^A 
\int d^3 x \int d^3 y \, |\vx-\vy| \{(\vz \times \vpi_j)_{\rm z} , 
\hat{e}_i (|\vx-\vxp_i|) \hat{e}_j (|\vy-\vxp_j|)\}
\, . \eqno(16)
$$
The one-body $(i=j)$ part of the convection-current contribution vanishes upon
integration, as does the second (tensor) term in the magnetization contribution.
Shifting the variables $\vx$ and $\vy$ by $\vxp_i$ leads to
$$
\delta_{\rm spin}^{(1)} = \sum_{i=1}^A \left( \langle r \rangle^{pp}_{(2)} 
\frac{\mup}{\muN}\, \hp_i + \langle r \rangle^{nn}_{(2)} \frac{\mun}{\muN}\,
\hn_i \right) \sigma_i^{\rm z} \, , \eqno(17)
$$
where
$$
\langle r \rangle^{pp}_{(2)} = \int d^3 x \int d^3 y \, \rho^p_{ch} (x) 
\rho^p_M (y) |\vx - \vy| = 1.086(12) {\, \rm fm} \, , \eqno(18a)
$$
and
$$
\langle r \rangle^{nn}_{(2)} = \int d^3 x \int d^3 y \, \rho^n_{ch} (x) 
\rho^n_M (y) |\vx - \vy| \,  \eqno(18b)
$$
determine the proton and neutron parts of the one-body current.  Note that the
quantities $\langle r \rangle^{pp}_{(2)}$ and $\langle r \rangle^{nn}_{(2)}$ are
the usual proton and neutron Zemach terms, and we have listed in Eqn.~(18a) the
value of the proton Zemach moment recently determined directly from the
electron-scattering data for the proton\cite{ingo04} (the neutron has not been
evaluated). In numerical work described below we will use simple forms for the
neutron and proton form factors:  a dipole form for the proton charge and
magnetic form factors and the neutron magnetic form factor $(F_D (q^2) =
\frac{1}{(1 + q^2/\beta^2)^2})$ and a modified Galster\cite{galster} form for
the neutron charge form factor $(F_G = \frac{\lambda \, q^2}{(1 +
q^2/\beta^2)^3})$.  To incorporate into our calculations the numerical value
given by Eqn.~(18a) we use $\beta = 4.0285$ fm$^{-1}$, which reproduces this
value for the dipole case (see Appendix A for moments and correlation functions
determined by this choice of form factors). The r.m.s. radius determined by this
$\beta$ is 0.86 fm, which is slightly smaller than the proton charge
radius\cite{ingo03} but slightly larger than its magnetic radius\cite{ingopc}
and thus represents an average value. The much smaller neutron moment, $\langle
r \rangle_{(2)}^{nn}$ (see Appendix A), can be adequately represented using this
value of $\beta$ and $\lambda = 0.0190$ fm$^2$, which determines the neutron
charge radius\cite{isotope}.  These numbers lead to $\langle r
\rangle_{(2)}^{nn} = -0.042$ fm, which we will use below. Because this value is
such a small fraction of the proton result, the uncertainty in the neutron value
plays no significant role.

Equation (17) is still a nuclear operator, and its expectation value depends on
the nucleus.  We begin with the deuteron (which has $T=0$) and this eliminates
the $\tau_3$ terms in $\hp_i$ and $\hn_i$.  The spin terms then sum to $(\vsig_1
+ \vsig_2)_{\rm z}$, which is not the total angular momentum (it lacks $\vL$,
the orbital angular momentum contribution).  The expectation value of $(\vsig_1
+ \vsig_2)_{\rm z}$ is $2 S_{\rm z}(1 - \frac{3}{2} P_D)$, where $S_{\rm z}$ is
the z-component of the deuteron total angular momentum operator and $P_D$ is the
amount of $D$-wave in the deuteron wave function (typically, slightly in excess
of 5.6\%).  In the state of maximum $S_{\rm z}$ this leads to
$$
\delta_d^{(1)} = (1 - \frac{3}{2} P_D) \left( \frac{\mup}{\mud}
\langle r \rangle^{pp}_{(2)} + \frac{\mun}{\mud}
\langle r \rangle^{nn}_{(2)} \right) \, , \eqno(19)
$$
which is completely dominated by the proton.  Note that the $D$-wave prefers to
anti-align with the spin, which leads to the reduction in Eqn.~(19).

A similar (though more complicated) analysis is possible for $^3$He and $^3$H
(see Appendix B).  The traditional (and very useful) decomposition of the
trinucleon wave function uses representations of spin-isospin symmetry (viz.,
SU(4)).  In addition to the somewhat larger D-state component $(P_D \sim 9\%)$,
the significant $S$-wave component comes in two distinct types:  the dominant
S-state ($P_S \sim 90\%$) with a completely antisymmetric spin-isospin wave
function and completely symmetric space wave function, and the mixed-symmetry
S$^{\prime}$-state $(P_{S^\prime} \sim 1\%)$.  The representations of SU(4) were
used long ago to decompose contributions to the trinucleon magnetic moments, and
this leads to\cite{pc,mag-mom}
$$\eqalignno{
\left \langle \sum_{i=1}^A \left( \frac{1 \pm \tau^3_i}{2} \right) \vsig_i 
\right \rangle =&
\vS \left( (1-2 P_D) \mp \tau_3 
(1 - \frac{4}{3} P_{S^{\prime}} - \frac{2}{3} P_D) \right) \, , &(20a)\cr
=& 2 \vS \left\{ \left[ \begin{array}{c} \gamma \\ \xi \end{array} \right]
\left( \frac{1+\tau_3}{2} \right) + 
\left[ \begin{array}{c} \xi \\ \gamma \end{array} \right]
\left( \frac{1-\tau_3}{2} \right) \right\} \, , &(20b) \cr}
$$
where
$$\eqalignno{
\gamma &= \frac{2}{3} \left( P_{S^{\prime}} - P_D \right)  \cong -0.06
\, , &(21a) \cr
\xi &= 1 - \frac{2}{3} P_{S^{\prime}} - \frac{4}{3} P_D \cong 0.86
\, , &(21b) \cr}
$$
specify the two sign cases in Eqns.~(20a) and (20b).  Note that $\tau_3$ is the
third component of total isospin and $\vS$ is the (total)
nuclear-angular-momentum operator. The first and second terms in Eqn.~(20b)
determine $^3$He and $^3$H, respectively, while upper and lower components refer
to $\pm$ in Eqn.~(20a). Finally we obtain
$$
\delta^{(1)}_{^3{\rm He}} =  \frac{\mup \gamma}
\muHe \langle r \rangle^{pp}_{(2)} + \frac{\mun \xi}
\muHe \langle r \rangle^{nn}_{(2)} \rightarrow 0 \, , \eqno(22a)
$$
and
$$
\delta^{(1)}_{^3{\rm H}} =  \frac{\mup \xi}
\muH \langle r \rangle^{pp}_{(2)} + \frac{\mun \gamma}
\muH \langle r \rangle^{nn}_{(2)} \rightarrow  \langle r \rangle^{pp}_{(2)}
\, , \eqno(22b)
$$
where the point-neutron, SU(4) symmetry limit (i.e., the S-state only, which is
indicated by the arrow) is discussed below.

This result is very easy to interpret.  In the dominant S-state (corresponding
to $\gamma = 0$ and $\xi = 1$) the two ``like'' nucleons (e.g., the protons in
$^3$He) have opposite and cancelling spins, while the ``unlike'' nucleon (e.g.,
the neutron in $^3$He) carries all of the spin and determines both the magnetic
moment and the single-nucleon contribution to the hyperfine structure (if we
ignore meson-exchange currents and the convection current). Ignoring those
currents leads to $\muHe = \gamma \mup + \xi \mun \rightarrow \mun$ and $\muH =
\xi \mup + \gamma \mun \rightarrow \mup$, where the arrow indicates the SU(4)
limit. The $^3$He single-nucleon contribution to hfs in this limit comes solely
from the tiny neutron contribution, while the corresponding $^3$H contribution
is simply $\langle r \rangle^{pp}_{(2)}$ and becomes identical to the free
proton Zemach moment.

The two-nucleon contributions are more complicated and are determined by
correlation functions.  In the deuteron these correlations must be between a
neutron and a proton and are of the types:  $e_p - \mun$, $e_n - \mup$, and
$e_n - e_p$ (convection current only).  In the trinucleons there are additional
types:  $e_p - \mup$, $e_n - \mun$, and $e_p - e_p$ (convection current only).
 As we noted earlier the correlations involving $e_n$ will be very small, and we
have ignored the tiny $e_n - e_n$ convection-current correlation in $^3$H.

The two-nucleon contributions contained in Eqns.~(15) and (16) can be
manipulated into simpler forms by shifting $\vx$ by $\vxp_i$ and $\vy$ by
$\vxp_j$, leading to
$$\eqalignno{
\delta_{\rm Low}^{\rm mag} =& \frac{1}{\muN}      \sum_{i \neq j}^A
\left( \vsig_j C_{i j} (x_{i j}) - \frac{1}{8} \bbC_{i j} (x_{i j})
(3 \vsig_j \cdot \hxij \hxij - \vsig_j)\right)_{\rm z} \, , &(23a) \cr
\delta_{\rm Low}^{\rm conv} =& \frac{3}{16 M \muN} \sum_{i \neq j}^A
\bar{C}_{i j} (x_{i j}) L_{i j}^{\rm z} \, , &(23b) \cr}
$$
where $\vL_{ij} = \vx_{ij} \times (\vpi_i - \vpi_j)$ and the three correlation
functions are defined by
$$\eqalignno{
C_{i j} (r)&    = \int d^3 x \int d^3 y \, \he_i (x) \hmu_j (y) |\vx -\vy +\vr|
                  \, , &(24a) \cr
          &    = \hp_i(\mup \hp_j + \mun \hn_j) C_{DD} (r)
                  + \hn_i(\mup \hp_j + \mun \hn_j) C_{DG} (r) \, , &(24b) \cr
\bC_{i j} (r)&  = \frac{1}{3 r}\frac{d\;}{dr} 
                  \int d^3 x \int d^3 y \, \he_i (x) \he_j (y) |\vx -\vy +\vr|^3
                   \, , &(24c) \cr
          &  = \hp_i \hp_j \bC_{DD} (r)
                  + (\hp_i \hn_j + \hn_i \hp_j ) \bC_{DG} (r) \, , &(24d) \cr
\bbC_{i j} (r)& = \frac{1}{3} \left( \frac{d^2\;}{dr^2} - \frac{1}{r}\frac{d\; }
                  {dr}\right) \int d^3 x \int d^3 y \, \he_i (x) \hmu_j (y) 
                  |\vx -\vy +\vr|^3 \, , &(24e) \cr
          &  = \hp_i(\mup \hp_j + \mun \hn_j) \, \bbC_{DD} (r) + 
             \hn_i(\mup \hp_j + \mun \hn_j) \, \bbC_{DG} (r) \, , &(24f) \cr}
$$
where in Eqns.~(24b,d,f) we have decomposed the charge and magnetic
distributions in terms of isospin projectors and radial functions $C_{DD} (r)$
and $C_{DG} (r)$. Explicit forms for these functions are given in Appendix A.
Note that the quantity in parenthesis in Eqns.~(24b,d) is the (dominant) spin
part of the magnetic moment operator, and that we ignore the contribution of two
neutron charge distributions in Eqn.~(24d). In the limit of no finite size each
of the three non-vanishing radial functions ($C_{DD}, \bC_{DD}, \bbC_{DD}$)
simply equals $r$.

The special case of the deuteron is easily dealt with.  With our assumptions
about the nucleon form factors there are only two distinct types of products
contained in $\hat{e}_i \hat{e}_j$  and $\hat{e}_i \hat{\mu}_j$; these are the
dipole-dipole form of $e_p-\mun$ correlation $(C_{DD})$ and the dipole-Galster
form of $e_n-e_p$ and $e_n-\mup$ correlation $(C_{DG})$. The conventional form
of the deuteron wave function (suppressing the spin and isospin wave functions)
is
$$
\psi_d = \left( u (r) + \frac{1}{\sqrt{8}} S_{1 2}(\hr) w (r) \right)
\left( \frac{1}{\sqrt{4 \pi} r}\right) \, , \eqno(25a)
$$
which leads to the useful relations
$$\eqalignno{
(4 \pi r^2) \psi_d^{\dagger} (\vr) \haf (\vsig_1 +\vsig_2 ) \psi_d (\vr) =& 
\vS A (r) + B (r) \left( 3 \hr \vS \cdot \hr -\vS \right) \, , &(25b) \cr
(4 \pi r^2) \psi_d^{\dagger} (\vr) \vL \psi_d (\vr)   =& 
\frac{3}{2}\vS D (r) -\frac{3}{4}D (r) \left( 3 \hr \vS \cdot \hr -\vS 
\right) \, , &(25c) \cr}
$$
using $\vL_{1 2} = 2\vL$ for the deuteron, while
$$\eqalignno{
A (r) =& u^2 (r) - \haf w^2 (r) \, ,                        &(26a) \cr
B (r) =& \frac{1}{\sqrt{2}} u (r) w (r) + \haf w^2 (r) \, , &(26b) \cr
D (r) =& w^2 (r) \, .                                       &(26c) \cr}
$$
This leads immediately to
$$\eqalignno{
\delta^{(2)}_d =& \int_0^{\infty} d r 
 \, A (r) \left( \frac{\mun}{\mud}\, C_{D D} (r) 
+ \frac{\mup}{\mud}\, C_{D G} (r) \right)
& \cr
                &-\frac{B (r)}{4}\, \left( \frac{\mun}{\mud}\, \bbC_{D D} (r) 
+ \frac{\mup}{\mud}\, \bbC_{D G} (r) \right) 
+ \frac{9}{4 \mud} D (r) \bC_{D G} (r) 
\, , &(27) \cr}
$$
where we have removed a factor of $1/2 M$ from the magnetic moments (i.e., they
are now given in units of nuclear magnetons).  In the limit of vanishing neutron
charge distribution and point protons, this expression becomes
$$
\delta^{(2)}_d \rightarrow \frac{\mun}{\mud} \int_0^{\infty} d r \, r 
\left( A (r) - \frac{B (r)}{4} \right) \, , \eqno(28)
$$ 
which is Low's expression\cite{d-th-1} for the complete deuteron finite-size
effect (in leading order). In the next Section we will refer to the integrals of
radial functions such as $A(r) C_{DD} (r)$, (i.e., including the numerical 
factors, but without the magnetic moments in Eqn.~(27)) as ``Low moments.'' 
The two Low moments for point-like nucleons are, therefore, $\int r A(r)$ and $-
\frac{1}{4} \int r B(r)$, as given in Eqn.~(28).

\section{Numerical Evaluation}

The proton hfs has been discussed in detail recently\cite{savely02,ingo04} and
we have nothing more to add. The recently evaluated proton Zemach moment is
listed in Table~II, and it leads to a $-$58.2(6) kHz contribution to the
hydrogen hfs, which equals $-$41.0(5) ppm. When added to the usual QED and 
recoil corrections\cite{eides01,savely02,ingo04} there is a 3.2(5) ppm
discrepancy with experiment, which can be attributed to hadronic polarization
and (possible) additional recoil corrections.

\begin{table}[ht]
\begin{center}
{\bf Table II}.
\caption{Neutron and proton Zemach moments and their single-nucleon contribution
to the deuterium 1s hyperfine structure.}
\vspace*{5mm}
\begin{tabular}{|ccl||cccc|}
\hline
\multicolumn{7}{|c|}{Nucleon Zemach Moments}\\ \hline
\multicolumn{3}{|c||}{Zemach moments}&
\multicolumn{4}{c|}{Deuteron Nucleon-Moment hfs}\\ \hline
proton & neutron & {} & proton & neutron & total& {}\\ \hline
1.086(12) & $-$0.042 & fm & $-$40.0 & $-$1.1 & $-$41.1 &kHz\\
\hline
\end{tabular}

\end{center}
\end{table}

We begin our discussion of the deuteron with the single-nucleon contribution
given by Eqn.~(19). Table~II lists the proton Zemach moment and the neutron
moment determined by our choice of form factors. The neutron result is only 4\%
of the proton value in magnitude, and the opposite sign reflects the fact that
the (overall neutral) neutron has negative charge at large distances that
balances positive charge at short distances. Using the value of $P_D \cong
5.67\%$ (corresponding to the AV18 potential model\cite{av18}) in Eqn.~(19)
leads to the nucleon-moment deuterium hfs contributions listed on the right side
of Table~II. The proton result differs from that in hydrogen by the factor of
$(1-\thaf P_D)$ and the statistical factors for the deuterium and hydrogen hfs.

\begin{table}[ht]
\begin{center}

{\bf Table III}.
\caption{Deuterium Low moments from various parts of the nuclear current. The
A-terms are the space-scalar contribution (first term in Eqn.~(23a)) from the
spin-magnetization current, the B-terms are the corresponding space-tensor terms
(second term in Eqn.~(23a)), while the D-term arises from the convection current
(Eqn.~(23b)).}
\vspace*{5mm}
\begin{tabular}{|cc|cccccl|}
\hline
\multicolumn{8}{|c|}{Deuteron Low Moments}\\ \hline
$A_{\rm pt}$ & $B_{\rm pt}$ & $A_{DD}$ & $A_{DG}$ & 
$B_{DD}$ & $B_{DG}$ & $D_{DG}$&{}\\ \hline
3.081&$-$0.115&3.271&$-$0.015&$-$0.126&0.001&$-$0.003&fm\\
\hline
\end{tabular}

\end{center}
\end{table}

The deuterium Low moments are listed in Table~III, and the resulting hfs is 
listed in Table~IV, both for point-like nucleons (only the proton charge 
contributes) and for nucleons with finite size. The moments themselves and the 
resulting hfs are defined in Eqn.~(28) for point-like nucleons and in Eqn.~(27)
for finite nucleons. It is obvious that d-waves and the neutron's charge 
distribution play a minor role. The proton charge distribution and the neutron 
magnetic distribution have a somewhat larger effect; $A_{D D}$ is larger than
$A_{\rm pt}$ by about 6\%.

\begin{table}[ht]
\begin{center}

{\bf Table IV}.  
\caption{Contributions to the deuterium 1s hyperfine structure from
the Low moments compiled in Table~III.}
\vspace*{5mm}
\begin{tabular}{|cc|cccccl|}
\hline
\multicolumn{8}{|c|}{Deuteron Low-Moment hfs}\\ \hline
$A_{\rm pt}$ & $B_{\rm pt}$ & $A_{DD}$ & $A_{DG}$ & 
$B_{DD}$ & $B_{DG}$ & $D_{DG}$&{}\\ \hline
84.9 & $-$3.2 & 90.2 & 0.6 & $-$3.5 & 0.0 & 0.0 & kHz\\
\hline
\end{tabular}

\end{center}
\end{table}

One can also compute the Zemach moment of the entire deuteron by constructing
the charge ($F_{\rm ch}$) and magnetic ($F_{\rm mag}$) form factors and using
the equivalent momentum-space version of the Zemach moment formula:
$$
\langle r \rangle_{(2)}  =  -\frac{4}{\pi} \int_0^{\infty} \frac{d q}{q^2} 
(F_{\rm ch} (q^2) F_{\rm mag} (q^2) - 1) \, . \eqno(29)
$$
Various contributions and limits are listed in Table~V. Results for point-like
nucleons are listed to the left and include the contributions from the s-wave
spin-magnetization current, the d-wave spin-magnetization current, and the
orbital (convection) current, followed by the total contribution. Including
identical dipole nucleon form factors for the proton's charge and the neutron's
magnetization densities (which multiplies both $F_{\rm ch} (q^2)$ and $F_{\rm
mag} (q^2)$ by $F_D (q^2)$ - see Eqn.~(A3)) leads to the rightmost result and
corresponds to an increase of about 10\%. The experimental result of 2.593(16)
fm was obtained directly from the electron-scattering data\cite{ingo04}, and
is approximately 2\% smaller (4 standard deviations) than our non-relativistic
calculation. This difference is the expected size of relativistic corrections
from MEC.

\begin{table}[ht]
\begin{center}

{\bf Table V}.  
\caption{Deuterium Zemach moments from various parts of the nuclear
current.}
\vspace*{5mm}
\begin{tabular}{|ccc|c||c||rl|}
\hline
\multicolumn{7}{|c|}{Deuteron Zemach Moments}\\ \hline
\multicolumn{4}{|c||}{Point N}&
\multicolumn{1}{|c||}{Finite N}&
\multicolumn{2}{|c|}{Experiment}\\ \hline
$L=0$ & $L=2$ & Orb & Zemach &{Zemach}&\multicolumn{2}{|c|}{Zemach}\\ \hline
2.324 & $-$0.035 & 0.094 & 2.383 & 2.656 & 2.593(16) &fm\\
\hline
\end{tabular}

\end{center}
\end{table}

\begin{table}[ht]
\begin{center}

{\bf Table VI}.  
\caption{Contributions to the deuterium hyperfine structure from 
one-nucleon and two-nucleon operators and their total.}
\vspace*{5mm}
\begin{tabular}{|ccc||cccl|}
\hline
\multicolumn{7}{|c|}{Deuteron hfs - Nucleon + Low Moments}\\ \hline
\multicolumn{3}{|c||}{Point N}& \multicolumn{4}{c|}{Finite N}\\ \hline
Nucleon & Low & Total & Nucleon & Low & Total &{}\\ \hline
0.0 & 81.8 & 81.8 & $-$41.1 & 87.3 & 46.2 & kHz\\
\hline
\end{tabular}

\end{center}
\end{table}

The one-body (nucleon Zemach) and two-body (Low) contributions to the total
deuterium hfs are listed in Table~VI. Because there is then no point-nucleon
contribution to the one-body part, the Low term is the sole contribution and
leads to a very large result. The finite-nucleon case has considerable
cancellation between the two, and totals only about half the size of the
point-nucleon limit. One can also break the total result down into deuteron
Zemach (elastic) terms plus polarization (inelastic) terms. This is indicated in
Table~VII. The polarizability term is more than twice the elastic (Zemach) term,
and reflects how easily a weakly bound system can be excited compared to a
system like the nucleon, which is difficult to excite. In the latter case the
polarization term is only about 10\% of the Zemach term.

\begin{table}[ht]
\begin{center}

{\bf Table VII}.  
\caption{Contributions to the deuterium hyperfine structure from 
elastic (Zemach) and inelastic (polarization) intermediate states.}
\vspace*{5mm}
\begin{tabular}{|ccc||cccl|}
\hline
\multicolumn{7}{|c|}{Deuteron hfs - Zemach + Polarization}\\ \hline
\multicolumn{3}{|c||}{Point N}& \multicolumn{4}{c|}{Finite N}\\ \hline
Zemach & Polar & Total & Zemach & Polar & Total &{}\\ \hline
$-$29.5 & 111.2 & 81.8 & $-$32.8 & 79.1 & 46.2 & kHz\\
\hline
\end{tabular}

\end{center}
\end{table}

The physics of the nuclear correction to the deuterium hfs is straightforward
and completely dominated by the proton Zemach moment and the $e_p - \mun$ Low
contribution. Nuclear structure plays little role except to fix the size of the
(radial) Low moment. The signs were fixed by the sign of the proton magnetic
moment (for the one-body term) and the neutron magnetic moment (for the two-body
term), and are opposite. If we incorporate the additional minus sign in
Eqn.~(11b), the naively expected sign of the one-body terms should be $-$, while
that of the Low contribution should be $+$, as we found in the deuterium case.
As we will see, however, nuclear structure can play an exceptional role in the
trinucleon, and these expectations are not fulfilled in two cases.

The required $^3$H and $^3$He matrix elements were calculated using wave
functions obtained from a Faddeev calculation\cite{jerry}. The
(second-generation) AV18\cite{av18} potential was used, together with an
additional TM$^{\prime}$ three-nucleon force\cite{TM} whose short-range cutoff
parameter had been adjusted for each case to provide the correct binding
energies. Individual one-body (labelled ``nucleon'') and two-body (labelled
``Low'') terms are tabulated together with their total in Table~VIII. Note that
the $^3$He case (which has proton number $Z=2$) is uniformly enhanced (compared
to the H cases) by a factor of $Z^3 = 8$ contained in $|\phi_n(0)|^2$ in
Eqn.~(11a). For the same reason the two protons in $^3$He effectively double
that Low moment. Taking those factors into account the $^3$He Low term becomes
comparable in size to that of the deuteron.

The (approximate) SU(4) symmetry that dominates light nuclei\cite{su4} provides
an explanation for the relative sizes of the entries in this table, as well as
the unexpected signs (see above) of the $^3$H one-body term and the $^3$He
two-body term. The two protons in $^3$He have their spins anti-aligned in the
SU(4) limit, and this cancellation leads to the small net result and unexpected
sign for the one-body part, which is determined by small components of the wave
function. The protons in $^1$H and $^3$H make comparable one-body contributions,
since the proton in $^3$H carries the entire spin in the SU(4) limit. For the
same reason the Low term in $^3$H is very small because the two neutron $e_p -
\mun$ terms largely cancel, since the proton carries all of the spin in the
SU(4) limit.

The neutron Zemach moment plays only a very small role in the one-body terms (as
we found for the deuteron) except for $^3$He, which has a greatly suppressed
proton contribution. The tensor term also is quite small. The convection current
terms are negligible for $^3$H, but the $e_p - e_p$ contribution in $^3$He is
approximately 5\% of the total.

\begin{table}[ht]
\begin{center}

{\bf Table VIII}.  
\caption{Contributions to the trinucleon hyperfine structure from one-nucleon
and
two-nucleon operators.}
\vspace*{5mm}
\begin{tabular}{|ccc||cccl|}
\hline
\multicolumn{7}{|c|}{Trinucleon hfs}\\ \hline
\multicolumn{3}{|c||}{$^3{\rm H}$}& \multicolumn{4}{c|}{$^3{\rm He}$}\\ \hline
Nucleon & Low & Total & Nucleon & Low & Total &{}\\ \hline
$-$50.6 & $-$9.6 & $-$60.1 & 14 & 1428 & 1442 & kHz\\
\hline
\end{tabular}

\end{center}
\end{table}

Our final results are listed in Table~IX. The first line of the table is the 
same as that in Table~I, showing the fractional difference of experiment and QED
theory in ppm. That fractional difference is recomputed in the second line when
the nuclear corrections are added to the theoretical result. In the proton case
the Zemach and recoil corrections slightly over-correct, but the overall result
is consistent with the expectation that the polarization corrections are
positive and must be less than 4 ppm\cite{faustov02,hughes83}. For the nuclear
cases the quality of our results must be considered quite good, given the size
of our hadronic expansion parameter. The deuterium case is particularly close to
experiment, and this is likely due to the small binding energy, which tends to
minimize relativistic corrections\cite{chi-pt}. The quality of our trinucleon
results range from very good in the $^3$H case ($\sim 3\%$ residue) to adequate
in the $^3$He case ($\sim 20\%$ residue). The large disparity in the two cases
is undoubtedly due to missing MEC, particularly the isovector ones. Even this
amount of missing strength is only slightly larger than our expansion parameter.

\begin{table}[htb]
\centering
{\bf Table IX}.
\caption{Difference between hyperfine experiments and hyperfine calculations
for the 1s-state of light hydrogenic atoms, expressed as parts per million 
of the Fermi energy. The first line is the difference with respect to the
QED calculations only, while the second line incorporates the hadronic
corrections (Zemach moment for the proton and nuclear corrections for the 
nuclei) calculated above.}
\vspace*{5mm}
\begin{tabular}{|l || c c c c|}
\multicolumn{5}{c}{$(E_{\rm hfs}^{\rm exp} - E_{\rm hfs}^{\rm Th})/E_{\rm F}
\, {\rm (ppm)}$} \\ \noalign{\smallskip} \hline
Theory        & H & $^2$H & $^3$H & $^3$He$^+$ \rule{0in}{2.5ex}\\ \hline \hline
QED only       & $-$33    & 138    & $-$38 & $-$212       \\ \hline
QED + hadronic & 3.2(5)   & $-$3.1 & 1.2     & $-$46          \\ \hline 
\end{tabular}
\end{table}

Previous work on this topic is quite
old\cite{d-th-1,d-th-0,t-th-1,t-th-2,he-th,d-t-he}, except for the
deuterium\cite{yulik} case. The older work relied on the Breit approximation for
the electron physics, which is sufficient only for the leading-order
corrections. It used an adiabatic treatment of the nuclear physics based on the
Bohr picture of the nuclear hyperfine anomaly, which is far more complex than
the treatment that we have presented. Uncalculated QED corrections and poorly
known fundamental constants (such as $\alpha$) led to estimates of nuclear
effects that were many tens of ppm in error. Although the nuclear physics at
that time was not adequate to perform more than qualitative treatments of the
trinucleons, the SU(4) mechanism was known and this allowed a qualitative
understanding. The only previous attempt to treat the three nuclei
simultaneously was in Ref.~\cite{d-t-he}. They found nuclear corrections of
about 200 ppm for deuterium, 20 ppm for $^3$H, and $-$175 ppm for $^3$He$^+$.
Except for the deuterium case (which involves significant cancellations) this
has to regarded as quite successful, given the knowledge available at that time.

\section{Conclusions}

We have performed a calculation of the nuclear part of the hfs for $^2$H,
$^3$H, and $^3$He$^+$, based on an expansion parameter adopted from $\chi$PT,
a unified nuclear model, and modern second-generation nuclear forces. This 
is the first such calculation, and the results are quite good. Details of
the results can be understood in terms of the approximate SU(4) symmetry
that dominates the structure of light nuclei. 

\section*{Acknowledgments}

The work of JLF was performed under the auspices of the U.\ S.\ Dept.\ of 
Energy, while the work of GLP was supported in part by the DOE.

\section{Appendix A}

The correlation functions that we require are built from the charge and magnetic
form factors for protons and neutrons.  The generic correlation function has the
form
$$
\rho_{(2)} (z) = \int d^3 x \, \rho_{\rm ch} (x) \, \rho_{\rm mag} (|\vx + \vz|)
\, , \eqno({\rm A}1)
$$
where the charge density $\rho_{\rm ch}(x)$ and magnetization density $\rho_{\rm
mag} (y)$ are normalized to one.  Although a variety of functional forms have
been proposed for these densities, few published forms have high accuracy over
the low-momentum-transfer region that is important for Zemach moments. 
Fortunately the proton's Zemach moment was recently determined to high accuracy,
and the neutron's is sufficiently small that any credible model should suffice.
For the neutron charge form factor we will assume a modified Galster 
form\cite{galster}
$$
F_G (q^2) = \frac{\lambda q^2}{(1+\frac{q^2}{\beta^2})^3} \, , \eqno({\rm A}2)
$$
with $\beta = 4.0285$ fm$^{-1}$ (determined below) and $\lambda$ = 0.0190 
fm$^2$. This is accurate enough\cite{isotope} for our purposes, both at low
values of $q^2$ and at moderate values of $q^2$. For the proton charge form
factor and the proton and neutron magnetic form factors we choose the tractable
and venerable dipole form
$$
F_D (q^2) = \frac{1}{(1+\frac{q^2}{\beta^2})^2} \, , \eqno({\rm A}3)
$$
which is a reasonable (but only moderately accurate) approximation.

These two forms inserted into Eqn.~({\rm A}1) generate the two correlation
functions that we require:  $\rho_{DD}(z)$ and $\rho_{D G}(z)$:
$$
4 \pi \rho_{D D} (z) = \frac{\beta^3}{48} \exp{(-\beta z)} 
(3 + 3(\beta z) + (\beta z)^2 ) \, , \eqno({\rm A}4)
$$
and
$$
4 \pi \rho_{D G} (z) = \frac{\lambda \beta^5}{384} \exp{(-\beta z)} 
(9 + 9(\beta z) + 2(\beta z)^2 -(\beta z)^3) \, . \eqno({\rm A}5)
$$
The first moment of these functions is the linear Zemach moment
$$
\langle r \rangle_{(2)}^{D D} = \int d^3 r \, r\,  \rho_{D D} (r)
= \int d^3 x \int d^3 y \, \rho_D (x)\, \rho_D (y) |\vx-\vy| = 
\frac{35}{8 \beta} \, , \eqno({\rm A}6)
$$
which we identify with the recently determined proton moment: 1.086(12)~fm. This
restricts $\beta$ to be 4.029(45) fm$^{-1}$, which we also use for the neutron.
The rms radius for a dipole with this value of $\beta$ is 0.86 fm, slightly
smaller than the proton's charge radius, but slightly larger than the magnetic
radius by a few percent, and this represents our level of accuracy (except for
the measured proton Zemach moment).

We can use these functions to determine the appropriate correlation functions
$$\eqalignno{
C_{D D} (r) &= \int d^3 x\, \rho_{D D} (x) |\vx +\vr|        & \cr
&= r + \frac{8}{(\beta^2 r)} -\frac{\exp{(-\beta r)}}{\beta}
\left(\frac{8}{(\beta r)} + \frac{29}{8} +\frac{5 (\beta r)}{8} 
+ \frac{(\beta r)^2}{24} \right) & \cr
& \rightarrow \frac{35}{8 \beta} + \cdots \, , &({\rm A}7) \cr}
$$
where the limiting form holds only for small ($\beta r$), and similarly
$$\eqalignno{
C_{D G} (r) &= \int d^3 x\, \rho_{D G} (x) |\vx +\vr|        & \cr
&= \lambda \beta \left( -\frac{2}{(\beta r)} + \frac{\exp{(-\beta r)}}{192}
\left( \frac{384}{(\beta r)} + 279 + 87(\beta r) +14(\beta r)^2 +(\beta r)^3
\right) \right) & \cr
& \rightarrow -\frac{35 \lambda \beta }{64} + \cdots \, . &({\rm A}8) \cr}
$$
The remaining functions that we require are determined by
$$\eqalignno{
C_{D D}^{\prime} (r) &= \int d^3 x\, \rho_{D D} (x) |\vx +\vr|^3  & \cr
&= r^3 + \frac{1}{\beta^3} \left(\frac{240}{(\beta r)} + 48(\beta r) 
- \exp{(-\beta r)} \left( \frac{240}{(\beta r)} + \frac{165}{2} 
+\frac{21 (\beta r)}{2} + \frac{(\beta r)^2}{2} \right) \right) & \cr
& \rightarrow \frac{315}{2 \beta^3} +\frac{35 r^2}{4 \beta} + \cdots \, ,
                                                           &({\rm A}9) \cr}
$$
and
$$\eqalignno{
&C_{D G}^{\prime} (r) = \int d^3 x\, \rho_{D G} (x) |\vx +\vr|^3      & \cr
&= \frac{\lambda}{\beta} \left( - 12(\beta r) -\frac{120}{(\beta r)} 
+ \frac{\exp{(-\beta r)}}{16}
\left( \frac{1920}{(\beta r)} + 975 + 207(\beta r) +22(\beta r)^2 +(\beta r)^3
\right) \right) & \cr
& \rightarrow -\frac{945 \lambda}{16 \beta} -\frac{35 \lambda \beta r^2}{32} + 
\cdots \, . &({\rm A}10) \cr}
$$
 From the former we obtain
$$\eqalignno{
&\bC_{D D} (r) = \frac{1}{3 r} \frac{\! d}{d r} C_{D D}^{\prime} (r)  & \cr
&= r + \frac{16}{(\beta^2 r)} -\frac{80}{\beta^4 r^3} +
\frac{\exp{(-\beta r)}}{\beta^4 r^3}
\left( 80 + 80(\beta r) + 24(\beta r)^2 + \frac{19 (\beta r)^3}{6} 
+ \frac{(\beta r)^4}{6} \right) & \cr
& \rightarrow \frac{35}{6 \beta} + \cdots \, , &({\rm A}11) \cr}
$$
and
$$\eqalignno{
&\bbC_{D D} (r) = \frac{1}{3}\left( \frac{d^2}{d r^2} 
-\frac{1}{r}\frac{\! d}{d r} \right) C_{D D}^{\prime} (r)  & \cr
&= r - \frac{16}{(\beta^2 r)} +\frac{240}{\beta^4 r^3} -
\frac{\exp{(-\beta r)}}{\beta^4 r^3}
\left( 240 + 240(\beta r) + 104(\beta r)^2 + 24(\beta r)^3 
+ 3(\beta r)^4 +\frac{(\beta r)^5}{6} \right) & \cr
& \rightarrow \frac{\beta r^2}{6} + \cdots \, , &({\rm A}12) \cr}
$$
and from the latter
$$\eqalignno{
&\bC_{D G} (r) = \frac{1}{3 r} \frac{\! d}{d r} C_{D G}^{\prime} (r)  & \cr
&= \lambda \beta \left(-\frac{4}{(\beta r)} +\frac{40}{(\beta r)^3} -
\frac{\exp{(-\beta r)}}{(\beta r)^3}
\left( 40 + 40(\beta r) + 16(\beta r)^2 + \frac{163 (\beta r)^3}{48} 
+ \frac{19(\beta r)^4}{48} +\frac{(\beta r)^5}{48}\right) \right) & \cr
& \rightarrow \lambda \beta \left (- \frac{35}{48} + \cdots \right ) \, ,
&({\rm A}13) \cr}
$$
and
$$\eqalignno{
\bbC_{D G} (r) &= \frac{1}{3}\left( \frac{d^2}{d r^2} 
-\frac{1}{r}\frac{\! d}{d r} \right) C_{D G}^{\prime} (r)  & \cr
&= \lambda \beta \left( \frac{4}{(\beta r)} -\frac{120}{(\beta r)^3} \right.
& \cr
&+ \left. \frac{\exp{(-\beta r)}}{(\beta r)^3}
\left( 120 + 120(\beta r) + 56(\beta r)^2 + 16(\beta r)^3
+ 3(\beta r)^4 + \frac{17(\beta r)^5}{48} + \frac{(\beta r)^6}{48} 
\right) \right) & \cr
& \rightarrow \lambda \beta \left( \frac{(\beta r)^2}{48} + \cdots \right) \, .
&({\rm A}14) \cr}
$$
Note that these functions have been normalized so that $C_{DD}$, $\bC_{DD}$,
and $\bbC_{DD}$ become $r$ in the limit of large $\beta$.

We resort to the simple zero-range approximation in order to make a rough
estimate of the effect of nucleon finite size on the dominant Low moment of the
deuteron. This approximation is most accurate for asymptotic (long-range)
quantities, but will substantially overestimate short-range effects. We ignore
the d-state and assume everywhere the asymptotic s-wave function $(N e^{-\kappa
r}/\sqrt{4\pi} r$, where $\kappa \simeq 0.235$ fm$^{-1})$. This leads to an
expansion parameter $x = 2 \kappa/\beta \simeq$ 0.115.  The matrix element of
$C_{DD}$ (relative to the point-nucleon value) is approximately $(1 - x^2
(\frac{13}{3} + 8 \ln{(x)} + \ldots)$, which produces an increase $\sim 18\%$
from nucleon finite size. This is too large by a factor of two compared to
detailed calculations, but shows that the finite-size effect is enhanced by the
large numerical coefficient of the logarithmic term beyond what is expected from
a ${\cal O}(x^2) \sim 1/\beta^2$ correction term.

\section{Appendix B}

Many features of the trinucleon systems, $^3$He and $^3$H, can be determined in
a semi-quantitative fashion (accurate at the $\sim$10\% level) by simplifying
the wave functions to the dominant component alone.  Wave function components
have traditionally been classified according to their combined spin-isospin
symmetry, determined by the generators of SU(4).  In this scheme the SU(4)
generators for a system of $A$ nucleons are determined by the intrinsic spins 
and isospins of the individual nucleons:
$$
\sigma^k        = \sum_{i=1}^A \sigma^k_i  \, , \eqno({\rm B}1)
$$
$$
\tau^{\alpha} = \sum_{i=1}^A \tau_i^{\alpha}  \, , \eqno({\rm B}2)
$$
and
$$
Y^{k \alpha}  = \sum_{i=1}^A \tau_i^{\alpha} \sigma_i^k  \, .  \eqno({\rm B}3)
$$
All wave function components are labelled by the (combined) intrinsic spin of
the three nucleons (${\cal S}$ = 1/2 or 3/2) and the total isospin ($T$ = 1/2 or
3/2). Wave function spin and isospin components are then determined by: (1) the
$[\,\overline{4}\,]$ or antisymmetric state (${\cal S}$ =1/2, $T$=1/2), which
combines with a completely symmetric space wave function to form the dominant
S-state; (2) the mixed-symmetry state, which can be separated into (${\cal S}$ =
1/2, $T$ = 1/2), $({\cal S}$ = 3/2, $T$=1/2), and $({\cal S}$ = 1/2, $T$ = 3/2)
components. The first term contributes to the small S$^{\prime}$-state, while
the second generates the D-state(s), and the third contributes only to tiny
isospin impurities.  The remaining spin-isospin representation has a tiny
$({\cal S}$ = 1/2, $T$ = 1/2) symmetric component (called the S$^{\prime
\prime}$-state, with a totally antisymmetric space wave function), and a $({\cal
S}$ = 3/2, $T$ = 3/2) D-wave isospin impurity.  The number of components in the
order discussed is (4) + (4 + 8 + 8) + (4 + 16) = 64, as expected. Ignoring the
tiny S$^{\prime \prime}$-state, very small isospin impurities, and the
negligible P-states, we can therefore write in an obvious but schematic notation
for the trinucleon wave functions
$$
\Psi_{\rm tri} = {\rm S} \oplus {\rm S}^{\prime} \oplus {\rm D} \, . 
\eqno({\rm B}4)
$$
It was shown many years ago\cite{pc} that expectation values of the SU(4) 
generators for the trinucleons have very simple forms.
$$
\langle \sum_{i=1}^A \vsig_i \rangle = 2 \vS (1 - 2 P_D) \, , \eqno({\rm B}5)
$$
and
$$
\langle \sum_{i=1}^A \vsig_i \tau_i^{\rm z} \rangle = - 2 \vS \tau_3 
\left (1 - \frac{4}{3} P_{\rm S^{\prime}} -\frac{2}{3} P_D \right) \, , 
\eqno({\rm B}6)
$$
where a spin-isospin expectation value of the nuclear spin and isospin operators
is still required.

As expected, the spatially symmetric S-state dominates the trinucleon ground
states $(P_S \sim 90\%)$ because it minimizes the kinetic energy.  The
mixed-symmetry S$^{\prime}$-state is much smaller $(P_{\rm S^{\prime}} \sim
1\%)$, while the very strong nuclear tensor force generates a relatively large
D-state component $(P_D \sim 9\%)$.  If one ignores the S$^{\prime}$-,
S$^{\prime \prime}$-, P-, and D-state components, the remaining S-state wave
function factorizes into a completely symmetric space wave function and a
completely antisymmetric spin-isospin wave function, which greatly facilitates
calculating matrix elements. The mixed spin-isospin generator $Y^{k\alpha}$ has
the very useful and simple property for the S-state
$$
Y^{k \alpha} | S \rangle = - 2 S^k \tau^{\alpha}| S \rangle  \, , 
\eqno({\rm B}7)
$$
which follows (except for the factor of $(-1)$) from the Wigner-Eckart theorem
and the properties of the $[\, \overline{4}\, ]$ state.


\begin{thebibliography}{999}
\bibitem{MEC} H.\ Arenh\"ovel, {\it Czech.\ J.\ Phys.\ }{\bf 43}, 259 (1993).
\bibitem{primer} J.\ L.\ Friar, {\it Few-Body Systems} {\bf 22}, 161 (1997).
\bibitem{eides01} M.\ I.\ Eides, H.\ Grotch, and V.\ A.\ Shelyuto,
             {\it Phys.\ Rep.} {\bf 63}, 342 (2001).
\bibitem{savely02} S.\ G.\ Karshenboim and V.\ G.\ Ivanov, 
             {\it Eur.\ Phys.\ J.\ D} {\bf 19}, 13 (2002); 
             {\it Phys.\ Lett.\ B} {\bf 524}, 259 (2002).
\bibitem{tjon} M.\ J.\ Zuilhof and J.\ A.\ Tjon, Phys.\ Rev.\ C {\bf 22}, 
            2369 (1980).
\bibitem{chi-pt} D.\ R.\ Phillips, G.\ Rupak, and M.\ J.\ Savage, 
            {Phys.\ Lett.\ B} {\bf 473}, 209 (2000).
\bibitem{franz} F.\ Gross, in {\it Modern Topics in Electron Scattering}, 
            ed.\ by B.\ Frois and I.\ Sick (World Scientific, Singapore,
            1991), p.\ 219. This lovely review thoroughly and clearly discusses
            relativistic effects in nuclear electromagnetic interactions.
\bibitem{edmonds} A.\ R.\ Edmonds, {\it Angular Momentum in Quantum Mechanics},
             (Princeton, 1960). Although nuclear physicists almost universally 
             denote the total nuclear angular momentum by $\vJ$, we have chosen
             to use $\vS$ instead. This more closely conforms to conventional
             usage in atomic physics. Nuclear physicists should be careful not 
             to confuse our $\vS$ with the simple sum of individual nucleon 
             spins.
\bibitem{BW} A.\ Bohr and V.\ F.\ Weisskopf, {\it Phys.\ Rev.} {\bf 77}, 94 
             (1950). This was the first detailed calculation of the combined 
             effect of the charge and magnetic nuclear densities on the 
             hyperfine structure in heavy atoms; it used a model for the nuclear
             densities. Zemach\cite{zemach56} investigated light atoms and used
             perturbation theory without assumptions about the form of the 
             nuclear densities.
\bibitem{zemach56} C.\ Zemach, {\it Phys.\ Rev.} {\bf 104}, 1771 (1956).
\bibitem{dhg} J.\ L.\ Friar, {\it Phys.\ Rev.\ C} {\bf 16}, 1540 (1977).
\bibitem{dipole} J.\ L.\ Friar and G.\ L.\ Payne, {\it Phys.\ Rev.\ C} 
             (Submitted).
\bibitem{d-th-1} F.\ Low, {\it Phys.\ Rev.} {\bf 77}, 361 (1950); F.\ E.\ Low
              and E.\ E.\ Salpeter, {\it Phys.\ Rev.} {\bf 83}, 478 (1951). This
              work calculated the correlation-function part of the nuclear hfs 
              for deuterium, as originally sketched by Bohr\cite{d-th-0}.
\bibitem{d-th-0} A.\ Bohr, {\it Phys.\ Rev.} {\bf 73}, 1109 (1948).
\bibitem{ingo04} J.\ L.\ Friar and I.\ Sick, {\it Phys.\ Lett.\ B} {\bf 579}, 
            285 (2004).
\bibitem{galster} S.\ Galster, {\em et al.}, {\it Nucl.\ Phys.} {\bf B32},
            221 (1971).
\bibitem{ingo03} I.~Sick, {\it Phys.\ Lett.\ B} {\bf 576}, 62 (2003).
\bibitem{ingopc} I.\ Sick, (Private Communication).
\bibitem{isotope} J.\ L.\ Friar, J.\ Martorell, and D.\ W.\ L.\ Sprung, 
            {\it Phys.\ Rev.\ A} {\bf 56}, 4579 (1997). This work contains a 
            discussion of the neutron's size.
\bibitem{pc} J.\ L.\ Friar, B.\ F.\ Gibson, and G.\ L.\ Payne, 
             {\it Ann.\ Rev.\ Nucl.\ Part.\ Sci.\ }{\bf  34}, 403 (1984). This
             review discusses the usefulness of those results.
\bibitem{mag-mom} E.\ L.\ Tomusiak, M.\ Kimura, J.\ L.\ Friar, B.\ F.\ Gibson, 
             G.\ L.\ Payne, and J.\ Dubach, {\it Phys.\ Rev.\ C} {\bf 32}, 
             2075 (1985). This reference tested an impulse approximation 
             formula (which is quite accurate, though not exact) for the 
             trinucleon magnetic moments that is based on an SU(4)
             decomposition: $\mu = \haf (\mup + \mun) [1- 2 P_D]
             -\haf (2 T_3) (\mup - \mun) [1 - \frac{4}{3} 
             P_{S^\prime} -\frac{2}{3}P_D] + \haf P_D [1+\frac{1}{3} 
             (2 T_3)]$, where the $P$ states have been ignored, and 
             $P_{S^\prime}$ and $P_D$ are the probabilities of the 
             S$^\prime$ and D states, respectively. The quantity $(2 T_3)$
             is +1 for $^3$He and \minus 1 for $^3$H.
\bibitem{jerry} J.\ L.\ Friar, G.\ L.\ Payne, V.\ G.\ J.\ Stoks, and 
             J.\ J.\ de Swart, {\it Phys.\ Lett.\ }{\bf B 311}, 4 (1993).
\bibitem{av18} R.\ B.\ Wiringa, V.\ G.\ J.\ Stoks, R.\ Schiavilla, 
              {\it Phys.\ Rev.\ C} {\bf 51}, 38 (1995).
\bibitem{TM} S.\ A.\ Coon, M.\ D.\ Scadron, P.\ C.\ McNamee, B.\ R.\ Barrett,
             D.\ W.\ E.\ Blatt, and B.\ H.\ J.\ McKellar, {\it Nucl.\ Phys.}
             {\bf A317}, 242 (1979); J.\ L.\ Friar, D.\ H\"uber, and  U.\ van 
             Kolck, {\it Phys.\ Rev.\ C} {\bf 59}, 53 (1999). The latter paper
             includes a $\chi$PT derivation of an improved Tucson-Melbourne 
             three-nucleon potential (originally derived in the former paper), 
             which is usually denoted TM$^{\prime}$. The latter can be viewed as
             a second-generation three-nucleon force.
\bibitem{su4} R.\ G.\ Sachs and J.\ Schwinger, {\it Phys.\ Rev.} {\bf 70}, 
              41 (1946).
\bibitem{faustov02} R.\ N.\ Faustov and A.\ P.\ Martynenko, 
              {\it Eur.\ Phys.\ C} {\bf 24}, 281 (2002).
\bibitem{hughes83} V.\ W.\ Hughes and J.\ Kuti, 
             {\it Ann.\ Rev.\ Nucl.\ Part.\ Sci.} {\bf 33}, 611 (1983).
\bibitem{t-th-1} R.\ Avery and R.\ G.\ Sachs, {\it Phys.\ Rev.} {\bf 74},
              1320 (1948).
\bibitem{t-th-2} E.\ N.\ Adams II, {\it Phys.\ Rev.} {\bf 81}, 
              1 (1951).
\bibitem{he-th} A.\ M.\ Sessler and H.\ M.\ Foley, {\it Phys.\ Rev.} {\bf 98}, 
              6 (1955).
\bibitem{d-t-he} D.\ A.\ Greenberg and H.\ M.\ Foley, {\it Phys.\ Rev.} 
              {\bf 120}, 1684 (1960).
\bibitem{yulik} I.\ B.\ Khriplovich and A.\ I.\ Milstein, 
              {\it Zh.\ Eksp.\ Teor.\ Fiz.} {\bf 125}, 205 (2004)
              [{\it JETP} {\bf 98}, 181 (2004)]; A.\ I.\ Mil{'shtein}, 
              I.\ B.\ Khriplovich, and S.\ S.\ Petrosyan, {\it J.\ Exp.\ Th.\ 
              Phys.} {\bf 82}, 616 (1996); {\it Phys.\ Lett.\ B} {\bf 366}, 
              13 (1996). These papers are noteworthy because they calculate the
              sub-leading-order correction to the deuterium hfs in zero-range
              approximation. We have calculated only the leading order here,
              but have done so in the spirit of traditional nuclear 
              calculations, while also calculating tritium and $^3$He$^+$.
\end{thebibliography}
\end{document}